\documentclass[a4paper]{jpconf}
\usepackage{graphicx}
\usepackage{cite}

\newcommand{\ie}{\textit{i.e}.}

\usepackage{todo}
\usepackage{hyperref}
\usepackage{tikz}
\usepackage[braket,qm]{qcircuit}
\usepackage{amsmath}
\usepackage{color,soul}
\usepackage[usestackEOL]{stackengine}

\edef\tmp{\the\baselineskip}
\setstackgap{L}{\tmp}

\begin{document}

\title{Towards practical Quantum Credit Risk Analysis}


\author{
    Emanuele Dri$^1$, Edoardo Giusto$^1$, Antonello Aita$^2$, Bartolomeo Montrucchio$^1$
}

\address{$^1$ DAUIN, Politecnico di Torino, Torino, Italy}

\address{$^2$ IBM Italia, IBM, Milano, Italy}
\ead{emanuele.dri@polito.it, edoardo.giusto@polito.it, antonello.aita@ibm.com, bartolomeo.montrucchio@polito.it}

\begin{abstract}
In recent years, a CRA (Credit Risk Analysis) quantum algorithm with a quadratic speedup over classical analogous methods has been introduced \cite{cra_journal}.
We propose a new variant of this quantum algorithm  with the intent of overcoming some of the most significant limitations (according to business domain experts) of this approach.
In particular, we describe a method to implement a more realistic and complex risk model for the default probability of each portfolio's asset, capable of taking into account multiple systemic risk factors.
In addition, we present a solution to increase the flexibility of one of the model's inputs, the Loss Given Default, removing the constraint to use integer values.
This specific improvement addresses the need to use real data coming from the financial sector in order to establish fair benchmarking protocols. 

Although these enhancements come at a cost in terms of circuit depth and width, they nevertheless show a path towards a more realistic software solution.
Recent progress in quantum technology shows that eventually, the increase in the number and reliability of qubits will allow for useful results and meaningful scales for the financial sector, also on real quantum hardware, paving the way for a concrete quantum advantage in the field. 

The paper also describes experiments conducted on simulators to test the circuit proposed and contains an assessment of the scalability of the approach presented. 
\end{abstract}


%

\section{Introduction}
\label{sec:introduction}

Quantum finance aims at exploiting the peculiar characteristics of quantum computing in order to solve a vast set of computational problems in the financial sector, outperforming classical counterparts \cite{quantum_finance}.
In the last few years, much attention has been dedicated to the possibility of achieving quantum advantage in Credit Risk Analysis (CRA) \cite{cra_journal}, an essential tool of risk management.
CRA is defined as the risk of loss arising from a debtor's insolvency \cite{basic_cra}. 
The majority of advanced CRA models for the Economic Capital (EC, i.e. the amount of capital that a company needs to ensure that it stays solvent given its risk profile) calculation make use of Monte Carlo methods.
These estimation techniques essentially rely on repeated random sampling to obtain numerical results \cite{var_mc}.

\begin{figure*}[!ht]%
   	\centering
    	\includegraphics[trim={1.5cm 13cm 1cm 10cm}, clip, width=1.0\textwidth]{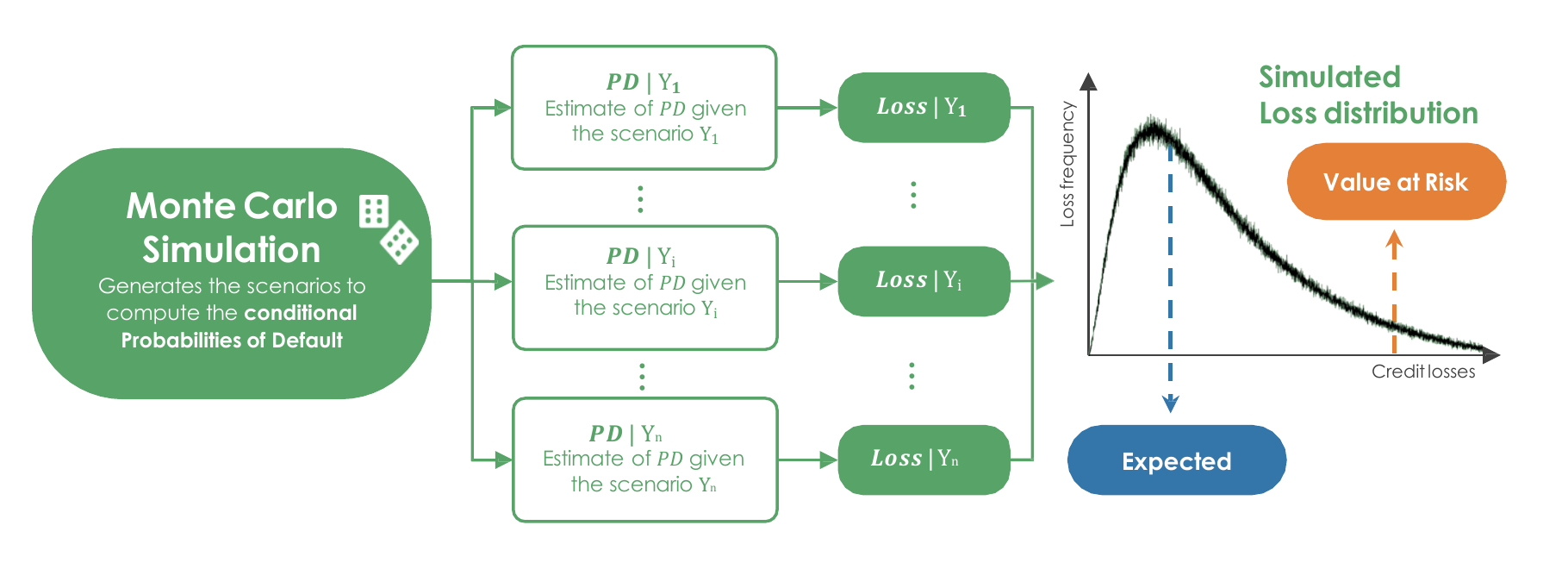}
        \caption{The Monte Carlo simulation process classically used for estimating the value at risk (VaR).}
        \label{fig:mc_cra}
\end{figure*}

A good example is Value at Risk (VaR), a statistic that quantifies how much a set of investments might lose (with a given probability) over a defined time frame \cite{var_def}.
This metric is broadly used for the assessment of the economic capital, but in most cases no closed-form solution currently exists for computing it \cite{closed_form}.
Therefore Monte Carlo simulations are a well-established choice to estimate its value and consequently evaluate the EC requirement \cite{var_mc}.
Figure \ref{fig:mc_cra} illustrates the simulation process.

The rate of convergence of the Monte Carlo method is $O(\sqrt{n})$ \cite{mc_rate}, where $n$ is the number of simulation paths.
The computational expensiveness of this approach derives from the nature of credit risk evaluation that falls in the category of rare-event simulation problems.

Moreover, another significant limitation of the classical Monte Carlo method is that it is possible at most to generate pseudo-random variables.
In fact, even when the available infrastructure of calculation used to accomplish this task is able to give in output millions of simulated scenarios, it’s advisable to preemptively fix a number generation threshold overtaking which the quality of the simulation is compromised by the systematic appearance of duplicates \cite{pseudorandom}.

This stimulated the search for new methods and solutions, exploiting new emerging fields such as quantum computing.
Quantum computers naturally overcome this restriction since they naturally generate true random samples due to the probabilistic nature of qubits \cite{true_random}.
In addition to this, the paradigmatic change operated by this technology offered new possibilities and favored innovative approaches able to take advantage of quantum technology.
In particular the quantum amplitude estimation (QAE) algorithm has been employed successfully in recent years to estimate the value at risk \cite{q_finance}, allowing for a quadratic speedup (on particularly limited scales) with respect to the classical Monte Carlo methodology.
This is achieved thanks to the increasing number of qubits in superposition used for the calculation.

Moreover, many of the complex custom models currently in use simulate default scenarios for predefined groups of counterparties \cite{cra_use}.
This is because executing the same simulation for each counterpart would be too demanding from the computational point of view in a classical computing architecture: even after an infrastructure scale-up, the gain in the simulation’s precision would not rebalance the execution time needed to produce the result.
Taking inspiration from the conventional Basel II model, the quantum approach estimates the measures of risk for every single counterpart \cite{cra_journal} and, once at scale, it could do so also when the considered number of the counterparties to evaluate is significantly higher, thanks to the quadratic speedup of the simulation process.

However, the existing quantum algorithm was designed to work with the already mentioned canonical framework Basel II \cite{cra_journal} based on an ASFR (asymptotic single-factor risk) model \cite{asfr}, which assumes that a borrower will default if the value of its assets were to fall below the value of its debts. 
Though this model helps to reserve an EC amount which suits every default scenario, it is intended to be a standard tool for CRA and therefore it is deliberately conservative \cite{basel_conservative}, rather than an optimal solution, especially for complex Credit Risk Portfolios.
As an example, Intesa Sanpaolo (Italy's largest bank by total assets \cite{top100}) uses a custom model that considers several risk factors instead of just one, since this refinement allows to reserve a more precise amount to cover potential losses \cite{multi-factor}. We modify the existing quantum algorithm to handle this increase in complexity preserving the advantages described above. 

Another possible enhancement of the aforementioned algorithm consists in finding a way to encode non-integer values for an input parameter to the problem, called Loss Given Default.
This would open the possibility to use real-world data in the quantum version of the model, and consequently compare the outcomes obtained fairly, using an actual benchmark. We also propose a solution for this specific issue, modifying the existing approach and exploring the costs in terms of needed quantum resources.

The rest of the paper is organized as follows: Section \ref{sec:qcra} briefly introduces the use of Quantum Amplitude Estimation for CRA, then Section \ref{sec:improvements} presents the different changes to the original algorithm in order to tackle those points described and not yet addressed by previous works, thus making the quantum architecture more adherent to real-world requirements. Finally, Section \ref{sec:results} presents the results of the simulation experiments done to validate the enhanced version of the original algorithm, and Section \ref{sec:conclusion} contains the relative conclusions, assessing what has been achieved and addressing the possibility of future improvements.

\section{Quantum Credit Risk Analysis}
\label{sec:qcra}
\subsection{Credit Risk Analysis with Quantum Amplitude Estimation}
Credit risk is typically quantified using three main measures: the probability of default (PD), the loss given default (LGD), and the economic capital ($E_{cap}$).

The PD is the probability that the debtor becomes insolvent.
The LGD corresponds to the estimated loss under the hypothesis of insolvency of the counterpart\footnote{Note that here LGD is to be understood as expressed in monetary units and not as a percentage of exposure (which instead is often done in risk management)}.
The expected loss, another commonly used measure of risk, instead depends on PD and LGD, since the higher these quantities are, the greater it will be.
The expected loss is obtained by multiplying the elementary risk measures mentioned above.
Given that this loss is an additive measure, the total expected loss for the $n$ assets that constitute a portfolio is equal to the sum of single exposure’s expected loss. 
\begin{equation}
\mathbb{E}[\mathcal{L}] = \sum_{k=1}^n PD_k\cdot LGD_k
\label{eq:expectd_loss}
\end{equation}

$E_{cap}$ is the third well-established metric which, as mentioned above, is defined as the amount of equity that a financial institution will hold to be able to cope with the risk of credit losses in its portfolio.
Given the distribution of losses, the economic capital is the value at risk (quantile of losses at a certain confidence level $\alpha$) minus the total expected loss.
\begin{equation}
E_{cap} = VaR_{\alpha}-\mathbb{E}[\mathcal{L}]
\label{eq:economic_capital}
\end{equation}

The total expected loss is subtracted from the VaR because the expected losses are usually already accounted for in a specific voice in the financial report of financial institutions.
Therefore, the economic capital does not refer to a concept of an average loss (like the expected loss does), but to an extreme or unexpected value of losses.

In \cite{cra_journal}, authors used the Quantum Amplitude Estimation (QAE) algorithm to estimate VaR with a quadratic speedup with respect to the classical Monte Carlo method. Recently a variant of QAE, called Iterative QAE (IQAE) has been proposed \cite{iqae}. This variant does not rely on Quantum Phase Estimation (QPE) but is only based on Grover’s Algorithm, which reduces the required number of qubits and gates, maintaining the quadratic speedup over classical alternatives. 

We refer the reader to \cite{qae}, \cite{q_finance}, and \cite{cra_journal} for additional information on QAE, while \cite{iqae} presents the improvements of the IQAE variant.
For the scope of this paper, we just give a general overview of quantum amplitude estimation from a high-level perspective. In order to apply QAE, the problem of interest has to be mapped to a quantum operator $\mathcal{A}$ acting on $n+1$ qubits such that
\begin{equation}
\mathcal{A}|0\rangle_{n+1}=\sqrt{1-a}\left|\psi_{0}\right\rangle_{n}|0\rangle+\sqrt{a}\left|\psi_{1}\right\rangle_{n}|1\rangle
\label{eq:amp_est}
\end{equation}
where $a \in[0,1]$ corresponds to the probability to measure $1$ in the last qubit, \ie, it is the (normalized) property of interest. Summarizing, quantum amplitude estimation allows the efficient estimation of $a$.

Therefore, to estimate the value at risk, authors of \cite{cra_journal} used quantum amplitude estimation in order to evaluate the cumulative distribution function of the total loss $\mathcal{L}$, constructing an operator $\mathcal{A}$ such that $a=\mathbb{P}[\mathcal{L} \leq x]$ for a given $x\geq 0$. Next, a bisection search is used to find the smallest $x_{\alpha}\geq 0$ for which $\mathbb{P}[\mathcal{L} \leq x_{\alpha}]\geq \alpha$; this implies that $x_{\alpha}=VaR_{\alpha}$.
Thus, when computing $VaR_{\alpha}$, the aim is to find the minimal threshold such that the estimated probability is larger than or equal to $\alpha$.

\subsection{Quantum Circuit for \texorpdfstring{$\mathcal{A}$}{TEXT}}

Three operators are needed in the original implementation to map the cumulative distribution function of the total loss to a quantum operator $\mathcal{A}$.
\begin{equation}
\mathcal{A=CSU}
\label{eq:a}
\end{equation}
The operator $\mathcal{U}$ loads the uncertainty model, the operator $\mathcal{S}$ computes the total loss into a quantum register with $n_S$ qubits and the operator $\mathcal{C}$ flips a target qubit if the total loss is less than or equal to a given level $x$ \cite{cra_journal}. 

For what concerns the default model, the scheme presented in \cite{cra_journal} is similar to the Basel II Internal Ratings-Based (IRB) approach, called the Gaussian conditional independence model \cite{gaussian_model},\cite{basel}.

According to this model, all losses can be expressed as $L_{k}=LGD_{k}\cdot X_{k}$ where $X_{k}\in\left\{ 0,1\right\}$ is a corresponding Bernoulli random variable. Here the probability that $X_{k} = 1$ corresponds to the probability of a default for asset $k$.

Following the Basel II approach, given a realization $z$ of a
latent random variable $\mathcal{Z}$ (also called systemic risk factor), the Bernoulli random variables $X_{k} \mid \mathcal{Z}=z$ are assumed independent, but their default probabilities $PD_k$ depend on $z$ while $\mathcal{Z}$ follows a standard normal distribution and
\begin{equation}
    PD_{k}(z)=F\left(\frac{F^{-1}\left(p_{k}^{0}\right)-\sqrt{\rho_{k}} z}{\sqrt{1-\rho_{k}}}\right)
    \label{eq:pd}
\end{equation}
where $p^{0}_k$ denotes the default probability for $z = 0$, $F$ is the cumulative distribution function of the standard normal distribution, and $\rho_{k} \in[0,1)$ determines the sensitivity of $X_k$ to $\mathcal{Z}$ \cite{cra_journal}.

\section{Proposed Improvements}
\label{sec:improvements}

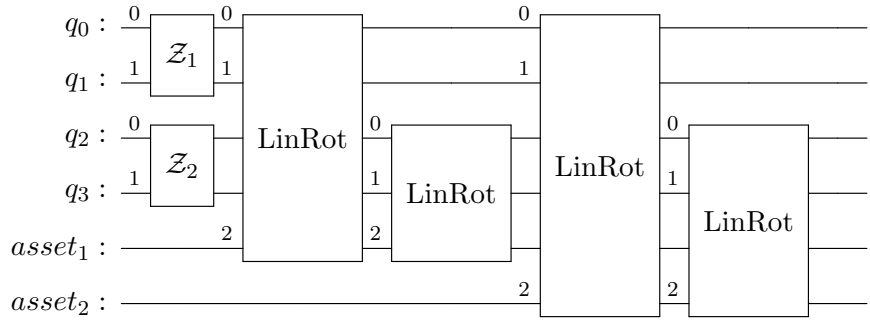
\begin{figure*}[!t]
\centering
\scalebox{1.0}{
\Qcircuit @C=1.0em @R=1.0em @!R { \\
	 	 \nghost{{q}_{0} :  } & \lstick{{q}_{0} :  } & \multigate{1}{\mathcal{Z}_1}_<<<{0} & \multigate{4}{\mathrm{LinRot}}_<<<{0} & \qw & \multigate{5}{\mathrm{LinRot}}_<<<{0} & \qw & \qw & \qw\\
	 	 \nghost{{q}_{1} :  } & \lstick{{q}_{1} :  } & \ghost{\mathcal{Z}_1}_<<<{1} & \ghost{\mathrm{LinRot}}_<<<{1} & \qw & \ghost{\mathrm{LinRot}}_<<<{1} & \qw & \qw & \qw\\
	 	 \nghost{{q}_{2} :  } & \lstick{{q}_{2} :  } & \multigate{1}{\mathcal{Z}_2}_<<<{0} & \ghost{\mathrm{LinRot}} & \multigate{2}{\mathrm{LinRot}}_<<<{0} & \ghost{\mathrm{LinRot}} & \multigate{3}{\mathrm{LinRot}}_<<<{0} & \qw & \qw\\
	 	 \nghost{{q}_{3} :  } & \lstick{{q}_{3} :  } & \ghost{\mathcal{Z}_2}_<<<{1} & \ghost{\mathrm{LinRot}} & \ghost{\mathrm{LinRot}}_<<<{1} & \ghost{\mathrm{LinRot}} & \ghost{\mathrm{LinRot}}_<<<{1} & \qw & \qw\\
	 	 \nghost{{asset}_{1} :  } & \lstick{{asset}_{1} :  } & \qw & \ghost{\mathrm{LinRot}}_<<<{2} & \ghost{\mathrm{LinRot}}_<<<{2} & \ghost{\mathrm{LinRot}} & \ghost{\mathrm{LinRot}} & \qw & \qw\\
	 	 \nghost{{asset}_{2} :  } & \lstick{{asset}_{2} :  } & \qw & \qw & \qw & \ghost{\mathrm{LinRot}}_<<<{2} & \ghost{\mathrm{LinRot}}_<<<{2} & \qw & \qw\\
\\
}}

\caption{Example of the multi-factor
version of the quantum circuit encoding the canonical uncertainty model, based on multiple rotations. The example has $K=2$ assets and $n_z=2$ \ie\ it uses two qubits to encode each normal standard distribution. Moreover there are two risk factors considered in the example ($R=2$) and as such two registers, for a total of four qubits, are needed to encode the two distributions.}
\label{fig:multi_rot}
\end{figure*}

This Section describes more in details the quantum algorithm under analysis, and illustrates the proposed improvements.

\subsection{Uncertainty model for Multiple Risk Factors}
\label{subsec:rot}

For the default events $\left\{X_{1}, \ldots, X_{K}\right\}$ it is possible to encode the $p^{0}_k$ of each asset in the state of a corresponding qubit by applying to qubit $k$ a Y-rotation $R_{Y}(\theta_{p_0}^k)$ with angle $\theta_{p_0}^{k}=2 \arcsin \left(\sqrt{p^{0}_k}\right)$.
Therefore the loading operator is
\begin{equation}
    \mathcal{U}=\otimes_{k=1}^{K} R_{Y}\left(\theta_{p_0}^{k}\right)
    \label{eq:u}
\end{equation}
To adjust $\mathcal{U}$ to include correlations between the default events, another register with $n_Z$ qubits to represent $\mathcal{Z}$ is added. The random variable $\mathcal{Z}$ follows a standard normal distribution. The original approach used a truncated and discretized approximation with $2^{n_Z}$ values, where an affine mapping $z_i=a_{z}i+b_z$ from $i\in\left\{ 0,\ldots,2^{n_Z}-1\right\}$ to the desired range of values of $\mathcal{Z}$ is used. 

$\mathcal{Z}$, being a discretized and truncated log-concave distribution, can be represented in a quantum register using an operator $\mathcal{U}_Z$. Such an operator is built from controlled rotations \cite{log-concave} which create a quantum state
\begin{equation}
    \sum_{i=0}^{2^{n_Z}-1} \sqrt{p_{i}}|i\rangle_{n_{Z}}
    \label{eq:u_z}
\end{equation}
where $p_i$ is the probability of observing outcome $i$ \cite{cra_journal}. The quantum register representing $\mathcal{Z}$ is then used to control the rotation
angles $\theta_{p}^{k}(z)=2 \arcsin \left(\sqrt{PD_{k}(z)}\right)$ that prepare the qubits representing the $X_k$.

However, as already stated in Section \ref{sec:introduction}, the Basel II single-factor model is deliberately conservative \cite{basel_conservative} and thus an extension of this approach is commonly used by large financial institutions to estimate default probabilities. 
The fundamental difference is the use of multiple systemic risk factors in place of just one, with the aim of directly attributing default correlations and, furthermore, default probabilities to the risk factors (while for the base model, given the realizations of the risk factors, defaults were uncorrelated).

Thus, the important advantage of this extended model relies upon the use of actual information about the point in time of the credit cycle. With this approach, uncertainties about the parameters which are needed for Value-at-Risk calculations in portfolio models can be reduced \cite{multi-factor}.

In the proposed implementation of the model described above, each risk factor (each $\mathcal{Z}_i$) still follows a standard normal distribution and has a weight $\alpha_i$ which financial institutions compute taking into account possible correlations effects among the different factors considered \cite{alphas}. Thus, the default probability effectively depends on a random variable $\mathcal{Y}$ which is a linear combination of the $R$ risk factors considered.
\begin{equation}
    \mathcal{Y} = \sum_{i=1}^R \alpha_i \mathcal{Z}_i
    \label{eq:y}
\end{equation}

From a practical perspective, we now have a model with multiple latent random variables whose realizations (appropriately combined) will determine the probability of default for each asset.
\begin{equation}
    PD_{k}(z)=F\left(\frac{F^{-1}\left(p_{k}^{0}\right)-\sqrt{\rho_{k}} \sum_{i=1}^{R}\alpha_i z_i}{\sqrt{1-\rho_{k}}}\right)
    \label{eq:pd_multi}
\end{equation}

To tackle such an increase in complexity this paper proposes two different alternatives to implement a quantum multi-factor version of the canonical uncertainty model.

\subsubsection{Multiple rotations}
the first proposal simply relies on multiple quantum registers for the encoding of the systemic risk factors (one register for each factor $\mathcal{Z}_i$). These correspond to multiple normal standard distributions, each one's realization controls one linear rotation for each asset (once weighted by the corresponding $\alpha_i$, using the slope of the rotation). As for the original algorithm, these rotations are then used to encode each asset's default probability in the amplitude of the qubit that represents that specific asset. Figure \ref{fig:multi_rot} shows the corresponding circuit.


This approach presents a limited overhead in terms of qubits required, in fact, the increase is only related to the extra qubits necessary for the new quantum registers that will represent the different risk factors. However, this approach requires a significant increase in the number of gates since each additional risk factor considered will require $K$ new controlled linear rotations.

\subsubsection{Single rotation}
this second version relies on one single quantum register encoding a random variable $\mathcal{N}$ which follows a multivariate normal distribution. A sum register is then used to add together the values taken by the normal distributions (corresponding to the marginal distributions of the multivariate one, with each marginal distribution representing a risk factor). The resulting value is then used to perform a single linear rotation for each asset, aimed at encoding its default probability in the target qubit.

In this case, the multivariate normal distribution is non-standard, since its covariance matrix is used to encode the $\alpha$ weights. This is done because it is not possible to encode the weights in the slope of the rotations, since here one single rotation per asset is performed, accounting for all the risk factors. This immediately shows the main limitation of this approach, which resides in the constraint of having the same $\alpha$ vector for all the assets.

The described solution allows for a reduction in terms of circuit depth with respect to the previous one since it relies on one single rotation for the encoding of the asset's default probability. However, such an approach generates an overhead in terms of needed qubits due to the presence of an extra sum register. On the other hand, such overhead becomes irrelevant in a setting with portfolios composed of thousands of assets. 

We invite the reader to see Section \ref{subsec:scal} for a more detailed assessment of the qubits and gates required by the various approaches.

\begin{figure}[t]
\centering
\scalebox{0.99}{
\Qcircuit @C=1.0em @R=1.0em @!R { \\
	 	\nghost{{q}_{0} :  } & \lstick{{q}_{0} :  } & \multigate{3}{\mathrm{P(X)}}_<<<{0} & \multigate{9}{\mathrm{adder}}_<<<{0} & \qw & \qw & \qw\\
	 	\nghost{{q}_{1} :  } & \lstick{{q}_{1} :  } & \ghost{\mathrm{P(X)}}_<<<{1} & \ghost{\mathrm{adder}}_<<<{1} & \qw & \qw & \qw\\
	 	\nghost{{q}_{2} :  } & \lstick{{q}_{2} :  } & \ghost{\mathrm{P(X)}}_<<<{2} & \ghost{\mathrm{adder}}_<<<{2} & \qw & \qw & \qw\\
	 	\nghost{{q}_{3} :  } & \lstick{{q}_{3} :  } & \ghost{\mathrm{P(X)}}_<<<{3} & \ghost{\mathrm{adder}}_<<<{3} & \qw & \qw & \qw\\
	 	\nghost{{q}_{4} :  } & \lstick{{q}_{4} :  } & \qw & \ghost{\mathrm{adder}}_<<<{4} & \multigate{6}{\mathrm{LinRot}}_<<<{0} & \multigate{7}{\mathrm{LinRot}}_<<<{0} & \qw\\
	 	\nghost{{q}_{5} :  } & \lstick{{q}_{5} :  } & \qw & \ghost{\mathrm{adder}}_<<<{5} & \ghost{\mathrm{LinRot}}_<<<{1} & \ghost{\mathrm{LinRot}}_<<<{1} & \qw\\
	 	\nghost{{q}_{6} :  } & \lstick{{q}_{6} :  } & \qw & \ghost{\mathrm{adder}}_<<<{6} & \ghost{\mathrm{LinRot}}_<<<{2} & \ghost{\mathrm{LinRot}}_<<<{2} & \qw\\
	 	\nghost{{q}_{7} :  } & \lstick{{q}_{7} :  } & \qw & \ghost{\mathrm{adder}}_<<<{7} & \ghost{\mathrm{LinRot}} & \ghost{\mathrm{LinRot}} & \qw\\
	 	\nghost{{q}_{8} :  } & \lstick{{q}_{8} :  } & \qw & \ghost{\mathrm{adder}}_<<<{8} & \ghost{\mathrm{LinRot}} & \ghost{\mathrm{LinRot}} & \qw\\
	 	\nghost{{q}_{9} :  } & \lstick{{q}_{9} :  } & \qw & \ghost{\mathrm{adder}}_<<<{9} & \ghost{\mathrm{LinRot}} & \ghost{\mathrm{LinRot}} & \qw\\
	 	\nghost{{asset}_{1} :  } & \lstick{{asset}_{1} :  } & \qw & \qw & \ghost{\mathrm{LinRot}}_<<<{3} & \ghost{\mathrm{LinRot}} & \qw\\
	 	\nghost{{asset}_{2} :  } & \lstick{{asset}_{2} :  } & \qw & \qw & \qw & \ghost{\mathrm{LinRot}}_<<<{3} & \qw\\
\\ }}
\caption{Example of the multi-factor
version of the quantum circuit encoding the canonical uncertainty model, based on one single rotation per asset. The example has the same parameters as the one depicted in Figure \ref{fig:multi_rot}.}
\label{fig:single_rot}
\end{figure}
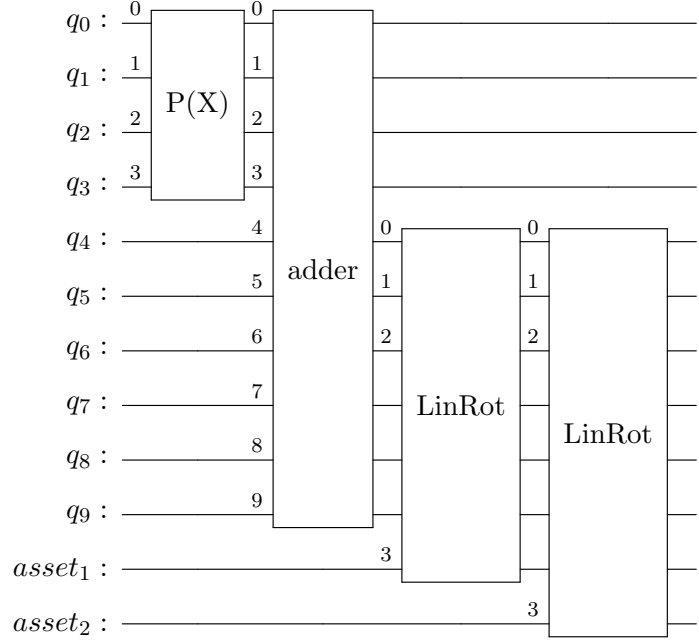

\subsection{Input flexibility avoiding the \texorpdfstring{$\mathcal{S}$}{TEXT} operator}
\label{subsec:flex}
Another limitation that characterizes available implementations of quantum credit risk algorithms regards the LGD parameters which can assume only integer values. This is due to the weighted sum register used to implement the operator $\mathcal{S}$ that computes the total loss. The formula below shows how $\mathcal{S}$ operates.
\begin{equation}
    \begin{aligned}
&\mathcal{S}:\left|x_{1}, \ldots, x_{K}\right\rangle_{K}|0\rangle_{n_{S}} \\
&\mapsto\left|x_{1}, \ldots, x_{K}\right\rangle_{K}\left|LGD_{1} x_{1}+\cdots+LGD_{K} x_{K}\right\rangle_{n_{S}}
\end{aligned}
\label{eq:s}
\end{equation}
Here $x_k\in \left\{0, 1\right\}$ denotes the possible realizations of $X_k$, while the loss given default of each asset is implemented using the weights of the \textit{WeightedAdder} register provided by Qiskit \cite{qiskit}, \cite{wadd} that are limited to assume integer values. Moreover, we need $n_{S}=\left\lfloor\log_{2}\left(LGD_{1}+\cdots+LGD_{K}\right)\right\rfloor+1$ qubits to represent in the second register all possible values of the sum of the losses given
default.

This constraint is especially limiting considering the small number of qubits currently available, in fact around 20 qubits would be needed only for the sum register, assuming 3 assets with values of $LGD \backsim \$100000$. Thus in order to be able to use realistic data as input for the quantum model, we propose a different version of the algorithm without the $\mathcal{S}$ operator.
This approach modifies the $\mathcal{C}$ operator using a circuit that implements a piecewise linear function $\hat{f}: \{ 0, ..., 2^n - 1 \} \rightarrow [0, 1]$ on qubit amplitudes \cite{laf}, \cite{laf2}, \cite{cra_journal}. 
\begin{equation}
    F|x\rangle|0\rangle = \sqrt{1 - \hat{f}(x)} |x\rangle|0\rangle + \sqrt{\hat{f}(x)}|x\rangle|1\rangle
    \label{eq:ampl_est}
\end{equation}
Where $|x\rangle$ is an $n$-qubit state.
This is done to allow for the operator to directly "read" the defaulted qubits from the X-register and associate the corresponding total loss. The objective qubit at this point is simply flipped if the total loss is less than or equal to the given level $x$ set by the current bisection search step.
Essentially the operator reads the X-register as a binary number, then the specific total loss (corresponding to the sum of the loss given default for that specific combination of defaulted qubits) associated with this number is used to make the comparison with $x$, flipping the objective qubit if needed. 

In the next section, this algorithm with the proposed improvements is applied to a small illustrative example using classical simulations of a quantum computer.

\section{Results}
\label{sec:results}

\begin{table}[t]
    \begin{center}
        \caption{Problem Parameters for the Two-Assets Example}
        \label{tab:table1}
        \begin{tabular}{lcccc}
            \hline \Longunderstack{asset \\ number} & \Longunderstack{loss given \\ default} & default prob. & sensitivity & \Longunderstack{risk factors \\ weights} \\
            $k$ & $LGD_{k}$ & $p_{k}^{0}$ & $\rho_{k}$ & $(\alpha_1,\alpha_2)_k$ \\
            \hline 1 & 1000.5 & $0.15$ & $0.1$ & $0.35, 0.2$\\
            2 & 2000.5 & $0.25$ & $0.05$ & $0.1, 0.25$\\
            \hline
        \end{tabular}
    \end{center}
\end{table}

\begin{figure}[!t]%
   	\centering
    	\includegraphics[width=0.48\textwidth]{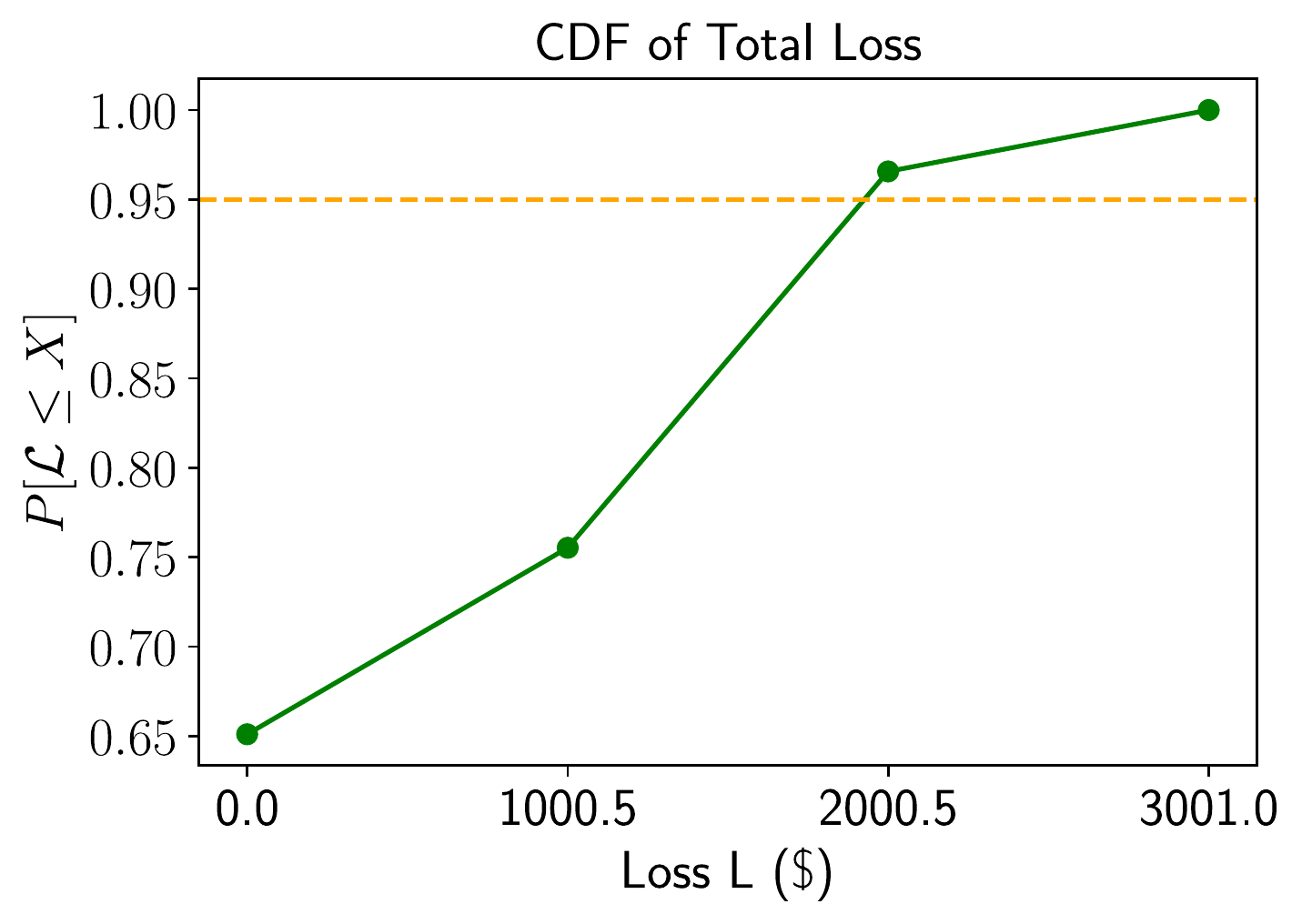}
        \caption{Cumulative distribution function of total loss $\mathcal{L}$ in green and target level of 95 percent in orange.}
        \label{fig:cdf}
\end{figure}

\begin{figure}[t]%
   	\centering
    	\includegraphics[width=0.48\textwidth]{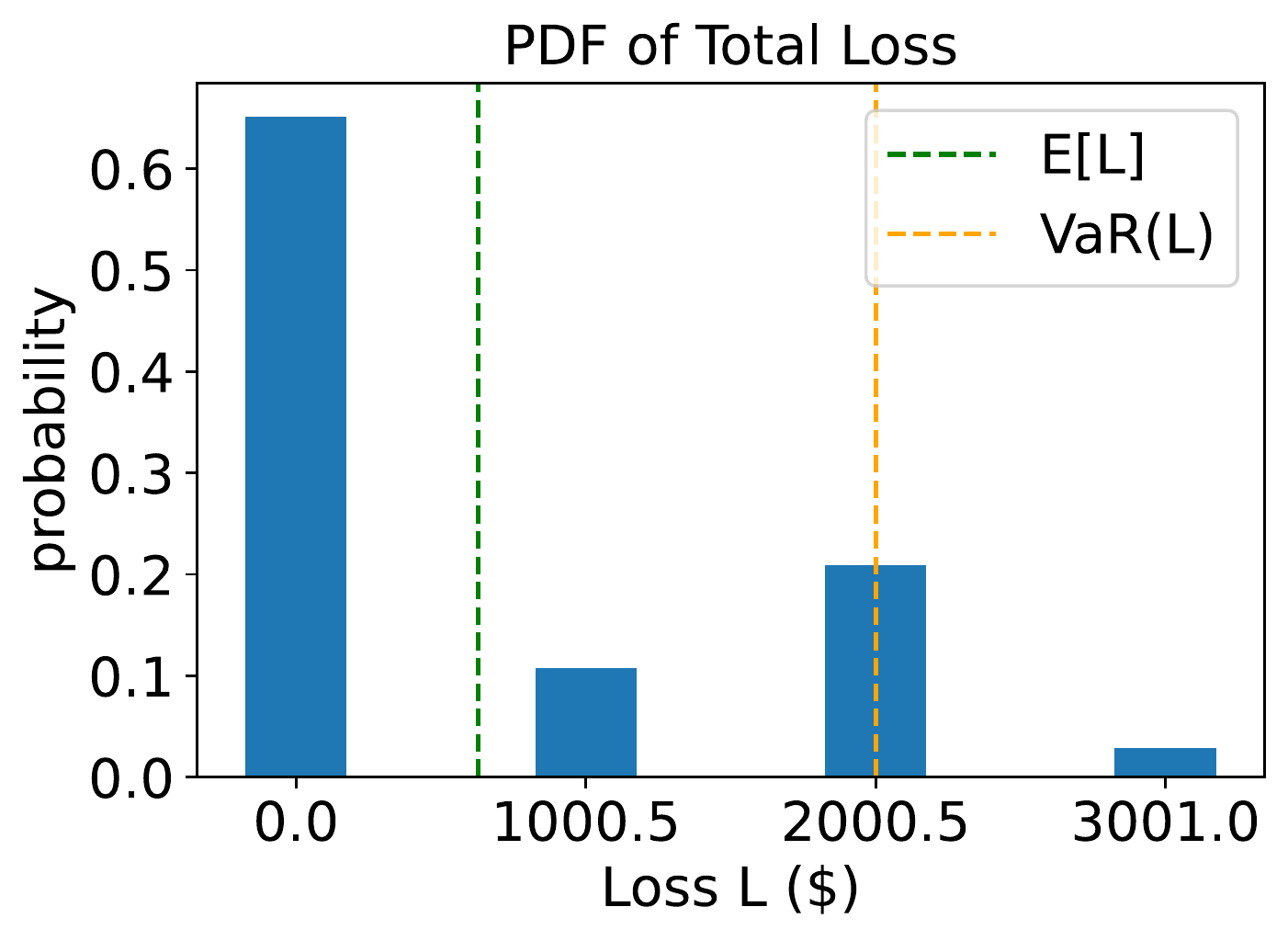}
        \caption{Probability distribution function of total loss. The green dashed line shows the expected loss while the orange dashed line shows the value at risk.}
        \label{fig:pdf}
\end{figure}

\begin{figure*}[ht]%
   	\centering
    	\includegraphics[width=1\textwidth]{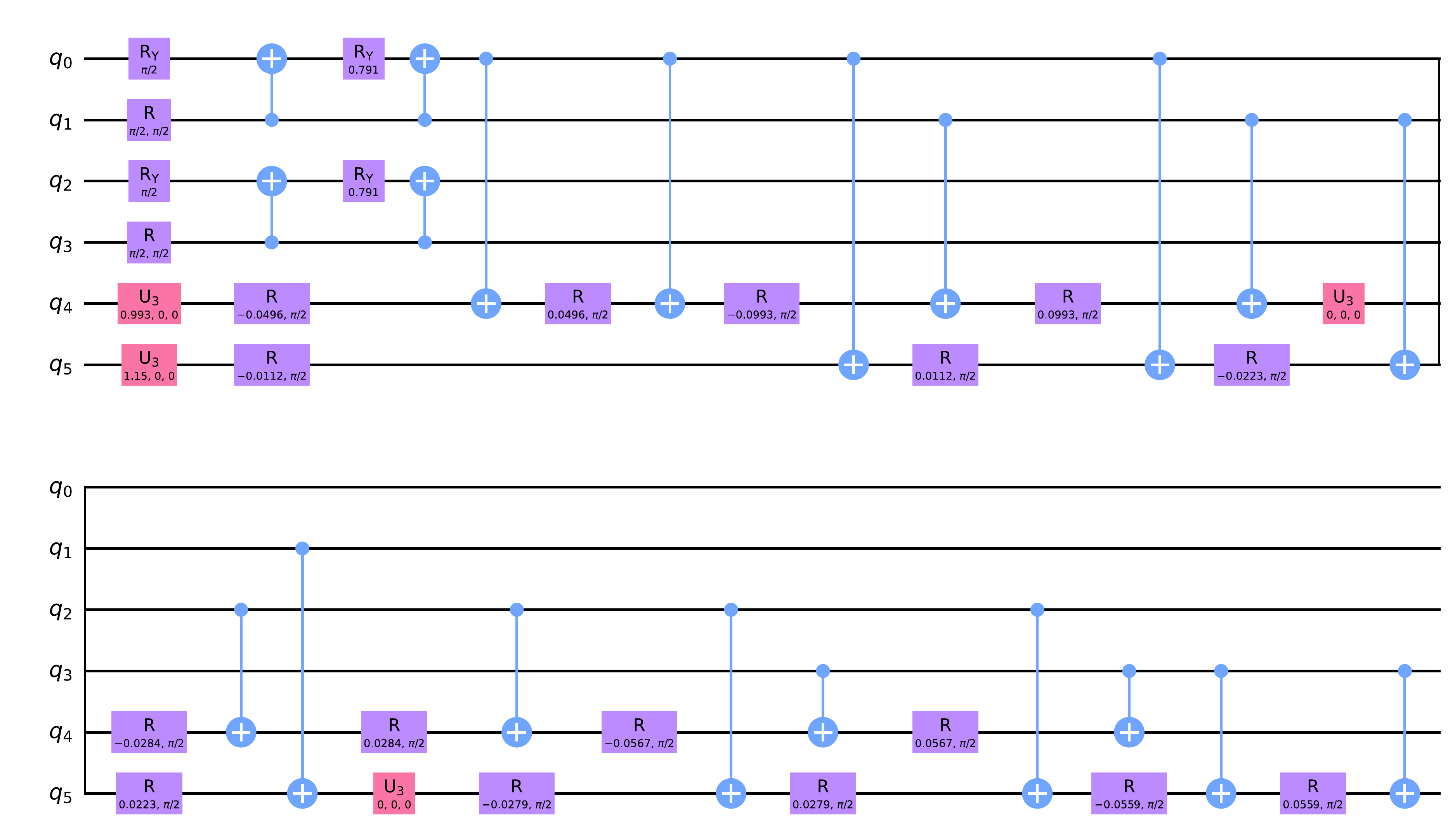}
        \caption{\centering Circuit for the multi-factor uncertainty model $\mathcal{U}$ encoded using the multiple-rotations variant: qubits $q_0$ and $q_1$ represent the Gaussian variable $\mathcal{Z}_1$, similarly $q_2$ and $q_3$ represent $\mathcal{Z}_2$, while $q_4$ and $q_5$ indicate whether the two assets default or not.}
        \label{fig:u}
\end{figure*}

This section contains the results of experiments conducted on a toy model designed to illustrate the proposed improvements. The experiments were conducted using the multiple-rotations scheme with $K=2$ assets and two systemic risk factors ($R=2$). The values for the different parameters are provided in Table \ref{tab:table1}.
The numeric values chosen for the LGD parameters have the specific intent of demonstrating the increased flexibility allowed by our approach with respect to the previous one.

Two qubits were used to model each one of the two latent random variables $\mathcal{Z}_1, \mathcal{Z}_2$, and no qubits were needed for the sum register since it is not necessary for the version of the algorithm we propose. 
The resulting loss distribution is shown in Figure \ref{fig:pdf}.
Moreover, Figure \ref{fig:cdf} shows the corresponding cumulative distribution function and the target level for the value at risk.

The simulation was carried out by supplying the circuit for $\mathcal{A}$ to the iterative amplitude estimation sub-routine implemented in Qiskit \cite{qiskit} and performing the bisection search, using the result to find the $VaR_{\alpha}$ (with $\alpha=0.95$). For the iterative quantum amplitude estimation, a target precision $\epsilon=0.002$ and a 99\% target confidence interval were set. This translated to an average of approximately 50000 quantum samples that the IQAE algorithm used to achieve the above mentioned precision and confidence. In total, this experiment required 9 qubits that were simulated on classical computers using the qasm simulator back-end provided by Qiskit \cite{qiskit}. 
The loading circuit for this experiment is shown in Figure \ref{fig:u}.

\subsection{Scalability}
\label{subsec:scal}
Although for small values of K (the number of assets considered) and R (the number of risk factors) the proposed multiple-rotations variant of the quantum model presents an advantage in terms of qubits required, thanks to the removal of the sum register and the use of the Iterative QAE, the scenario changes when the algorithm scales to a realistic setting with thousands of assets and tens of factors.
In fact, at such scaling the overhead derived from the presence of the sum register becomes negligible, since the number of qubits it requires scales logarithmically as 
$$O(\log_2 (\sum_{i=1}^K LGD_i))$$ 

Instead, using the Qiskit \textit{LinearAmplitudeFunction} register \cite{laf}, which is necessary to allow the input flexibility described in Section \ref{subsec:flex}, requires one additional qubit for each asset taken into account, therefore doubling the increase in terms of qubits that each additional asset entails. From a practical perspective, this translates into an increase in the width of the circuit with respect to the number of assets $K$ that goes as $O(2K)$ instead of $O(K)$ which was the rate for the implementation in \cite{cra_journal}.

Moreover, for both of the proposed variants, the increase in the number of factors comports a linear increment in the number of the required qubits which is proportional to $n_Z$.

For what concerns the circuit depth, we refer the reader to \cite{q_finance} and especially \cite{cra_journal} which contains an exhaustive analysis dedicated to the number of gates required for the original implementation.
For our implementations, it must be highlighted the increase in terms of needed gates due to the use of the \textit{LinearAmplitudeFunction} register. In fact, this circuit uses controlled linear rotations and comparator registers to implement the piecewise linear function on qubit amplitudes \cite{laf}. The number of such registers (and thus of the required gates) increases as $O(2^K)$. 

The Iterative QAE routine used to estimate the desired output does not introduce an increase in the number of required qubits since it does not require performing quantum phase estimation and still provides convergence proofs (which are instead missing for many of the other variants of the original QAE algorithm, such as the one in \cite{mle_qae}). Still, IQAE inevitably increases the depth of the circuit proportionally to the number of quantum samples needed to achieve the target precision. We refer the reader to \cite{iqae} for more details regarding the above-mentioned routine and its convergence rate.

Regarding the two different possible variants used to implement the multi-factor risk model, the overhead for what concerns the number of needed qubits is already described in Section \ref{subsec:rot}. Regarding the number of required gates, and therefore the depth of the circuit, the most expensive approach is the multiple-rotations one (which however is also the more flexible) since it requires a number of controlled rotations that grows linearly with the number of assets but also with the number of risk factors, while for the single-rotation approach it scales as $O(K)$ and is thus independent of the number of risk factors.

All these increments in terms of needed quantum resources have to be expected given the increased complexity and flexibility of the introduced circuits. 

\section{Conclusion}
\label{sec:conclusion}

In this paper, we presented several solutions addressing some of the shortcomings that currently prevent the Quantum Credit Risk Analysis algorithm from being an effective tool, ready for the coming improvements in terms of reliability and availability of qubits and quantum architectures in general.
We have demonstrated the capabilities of our approach simulating it, presented the results of such simulation and analyzed the scaling of the enhanced quantum algorithm.

From this analysis, it emerges how further improvements are required in terms of scalability for the proposed measures, since all of them require significantly more gates and qubits at scale than the previous implementation does.

The authors are aware of the importance of obtaining experimental results on real hardware. However, the test conducted showed how currently, the effects of noise completely destroy the output when the execution is performed on real quantum architectures. In particular, we need a significant improvement for what concerns qubits' coherence time, in order to observe meaningful results even outside ideal simulations.

Future works should also concentrate on creating a benchmark for the quantum model in order to have a fair comparison with classical production algorithms currently used by financial institutions. This is now possible thanks to
\begin{itemize}
    \item the ability to use real-world data, with the increase in input flexibility granted by the improvements proposed in Section \ref{subsec:flex}. In particular, this can be achieved since our architecture can take as input non-integer values for the LGD vector.
    \item the new and more realistic uncertainty model with multiple risk factors, which is the one commonly used by big entities in the financial sector \cite{cra_use}, \cite{cra_use_emerging}.
\end{itemize}





\section*{Acknowledgments}

The authors would like to thank Roberto Ugoccioni, Davide Corbelletto and Davide Ricossa from Intesa Sanpaolo for the inestimable insights they provided regarding the use case as well as on the improvements needed from a business perspective, and the team of IBM Italy for their valuable help and dedication to the development of the premises for this work. Finally many thanks to Stefan Woerner whose knowledge was instrumental for the single-rotation multi-factor variant of the uncertainty model.

\section*{References}

\bibliographystyle{IEEEtran}
\bibliography{refs}

\clearpage


\end{document}